\documentclass[%
reprint,
superscriptaddress,
amsmath,amssymb,
]{revtex4-2}

\usepackage{graphicx}
\usepackage{dcolumn}
\usepackage{float}
\usepackage{bm}

\begin{document}
\title{Zero-field magnetic structure of the antiferromagnetic metal EuSnP}
\author{Mizuki Urai}
\email{urai@issp.u-tokyo.ac.jp}
\affiliation{The Institute for Solid State Physics, University of Tokyo, Kashiwa, Chiba 277-8581, Japan.}
\author{Hiraku Saito}
\affiliation{The Institute for Solid State Physics, University of Tokyo, Kashiwa, Chiba 277-8581, Japan.}
\author{Daisuke Nishio-Hamane}
\affiliation{The Institute for Solid State Physics, University of Tokyo, Kashiwa, Chiba 277-8581, Japan.}
\author{Taro Nakajima}
\affiliation{The Institute for Solid State Physics, University of Tokyo, Kashiwa, Chiba 277-8581, Japan.}
\affiliation{RIKEN Center for Emergent Matter Science, Wako 351-0198, Japan.}
\author{Rina Takagi}
\email{rina.takagi@issp.u-tokyo.ac.jp}
\affiliation{The Institute for Solid State Physics, University of Tokyo, Kashiwa, Chiba 277-8581, Japan.}

\date{\today}

\begin{abstract}
We investigated the zero-field magnetic structure of the antiferromagnetic metal EuSnP through single-crystal neutron diffraction experiments. The magnetic propagation vector in the magnetically ordered phase was determined to be $(0,0,1/2)$, indicating that commensurate antiferromagnetic ordering is realized in this material. The detailed magnetic structure analysis revealed a collinear A-type antiferromagnetic structure with ferromagnetic alignment of Eu$^{2+}$ moments within the Eu--P layer.
The present findings provide a microscopic basis for understanding the intriguing magnetic behaviors of EuSnP, as exemplified by the multiple metamagnetic transitions in the antiferromagnetic state and the strong pressure dependence of the magnetic transition temperature.
\end{abstract}

\maketitle

\section{Introduction}
Antiferromagnetic metals hosting localized magnetic moments and itinerant electrons offer a platform for exploring the interplay between magnetism and electronic properties. 
Recent studies have revealed unconventional electromagnetic responses in systems with diverse magnetic structures, including noncollinear~\cite{Nature-2015-Nakatsuji}, noncoplanar~\cite{NCommun-2018-Ghimire,NPhys-2023-Takagi}, and even collinear configurations~\cite{NElectron-2022-Feng}, despite the absence of a large net magnetization.
These findings shed light on the importance of clarifying the magnetic structures of antiferromagnetic metals for understanding the origin of their characteristic behaviors.

\begin{figure}
	\includegraphics{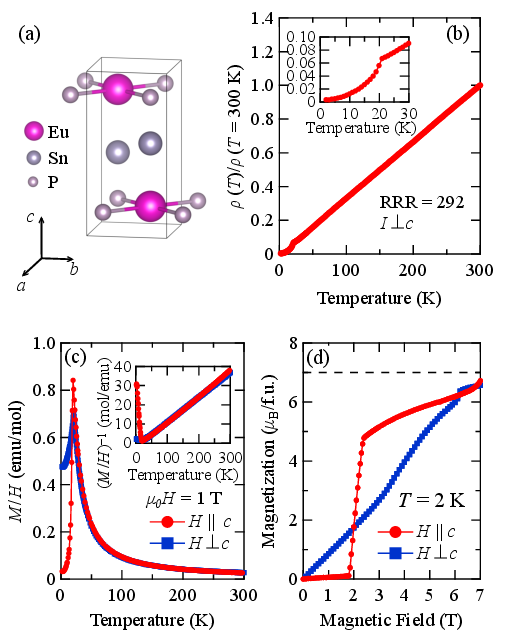}
	\caption{\label{fig:basic_properties} (a) Crystal structure of EuSnP~\cite{JAllyoysCompd-2002-Payne}, drawn using visualizing software VESTA~\cite{VESTA}. (b) Temperature dependence of resistivity normalized at 300 K. (Inset) Enlarged plot below 30 K. (c) Magnetic susceptibility measured at 1 T. (Inset) Inverse of magnetic susceptibility. (d) Magnetization curves at 2 K.}
\end{figure}

EuSnP is a system in which localized divalent Eu$^{2+}$ moments with the $4f^{7}$ electronic configuration ($L=0, S=7/2$, and $J=7/2$ where $L$, $S$ and $J$ denote the orbital, spin and total angular momenta, respectively) coexist with Sn and P conduction electrons in a tetragonal crystal structure [Fig.~\ref{fig:basic_properties}(a)].
EuSnP undergoes an antiferromagnetic transition at $T_{\mathrm{N}} \sim$ 21~K~\cite{JAllyoysCompd-2002-Payne}.
In the magnetically ordered state, multiple metamagnetic transitions are observed under magnetic fields applied both parallel and perpendicular to the $c$ axis~\cite{PhysicaB-2006-Fujiwara,JPSCP-2020-Iha,SciRep-2025-Podgorska,JPSJ-2025-Miyake}, concomitant with anomalies in the magnetoresistance and Hall resistivity~\cite{JPSJ-2025-Miyake}.
Another notable feature is that the antiferromagnetic state remains robust under applied pressure~\cite{DaltonTrans-2019-Gui,JPSCP-2020-Iha,JPSJ-2025-Miyake}, where $T_{\mathrm{N}}$, as identified by a kink in the resistivity, increases nearly linearly with pressure and reaches 140 K at 13.5 GPa~\cite{JPSJ-2025-Miyake}.
These field- and pressure-induced responses suggest the presence of nontrivial magnetic interactions in EuSnP, whose microscopic origin remains unclear.
Although possible magnetic structures have been proposed based on M\"{o}ssbauer spectroscopy~\cite{SciRep-2025-Podgorska}, the detailed magnetic structure has yet to be directly determined.

In the present study, we elucidate the zero-field magnetic structure of EuSnP using single-crystal neutron diffraction. The identified collinear A-type antiferromagnetic order, in which Eu$^{2+}$ moments are aligned ferromagnetically within the Eu–P layers and antiferromagnetically along the stacking direction, provides a basis for understanding the nontrivial field- and pressure-induced properties of EuSnP.

\section{Experimental Methods}
Single crystals of EuSnP were synthesized by the Sn self flux method based on Ref.~\cite{JPSCP-2020-Iha}. 
The crystal structure was confirmed by single-crystal X-ray diffraction, and the crystal orientation was determined by X-ray Laue photography. The chemical composition of a piece of grown crystals was analyzed by scanning electron microscopy with energy dispersive X-ray spectroscopy, yielding a composition of Eu: Sn: P = $0.99:1.08:0.94$.
Charge transport properties and magnetic susceptibility were measured for the evaluation of the samples using a Physical Property Measurement System (PPMS, Quantum Design Inc.) and Magnetic Property Measurement Systems (MPMS-5S and MPMS-XL, Quantum Design Inc.), respectively.
Single-crystal neutron diffraction measurements were carried out using the triple-axis neutron spectrometer PONTA installed at the 5G beamline in Japan Research Reactor 3 (JRR-3)~\cite{JPSJ-2024-PONTA}.
The incident neutron beam, with an energy of $E_{i}=34.069$ meV, was obtained by monochromatizing an unpolarized polychromatic beam from the reactor using a pyrolytic graphite monochromator.
The sample, which had a large flat surface along the $ab$ plane and dimensions of 1.24 mm $\times$ 2 mm $\times$ 74 $\mu$m, was probed in the $(h,h,l)$ and $(h,0,l)$ scattering planes.
For the magnetic structure analysis, $\theta$--$2\theta$ scans were performed at 2~K for each magnetic reflection.
The integrated intensities at 2 K, after subtraction of the 25~K ($> T_{\mathrm{N}}$) background and corrections for neutron absorption and the Lorentz factor, were used to obtain the observed magnetic structure factor, $|F^{\mathrm{M}}_{\mathrm{obs}}|$.
The scale factor for the magnetic structure analysis was determined by a least-squares comparison between the observed and calculated nuclear structure factors, $|F^{\mathrm{N}}_{\mathrm{obs}}|$ and $|F^{\mathrm{N}}_{\mathrm{cal}}|$, respectively, as described in Appendix~\ref{appendix:structure_analysis}.

\section{Results}
\subsection{Charge-transport and Magnetic properties}
The temperature dependence of the electric resistivity of a EuSnP single crystal synthesized in this work is shown in  Fig.~\ref{fig:basic_properties}(b).
The resistivity decreases linearly with temperature upon cooling and exhibits a kink at approximately 21~K, coinciding with  $T_{\mathrm{N}}$ determined from the magnetic susceptibility [Fig.~\ref{fig:basic_properties}(c)].
The residual resistivity ratio (RRR), calculated as the resistivity at 2 K normalized by that at 300 K, was 292, indicating the high quality of the sample.
The magnetic susceptibility, $M/H$, shows easy-axis behavior along the $c$ axis, as shown in Fig.~\ref{fig:basic_properties}(c), consistent with the previous reports~\cite{JAllyoysCompd-2002-Payne,PhysicaB-2006-Fujiwara,JPSCP-2020-Iha,PRMater-2024-Sprague,SciRep-2025-Podgorska}.
The inverse of the magnetic susceptibility is well fitted above 50~K by the Curie--Weiss relation given by  $(M/H)^{-1} = (T-{\Theta}_{\mathrm{W}})/C$, where $T$, ${\Theta}_{\mathrm{W}}$ and $C$ represent temperature, the Weiss temperature and the Curie constant, respectively, yielding $\Theta_{\mathrm{W}} = 16.0$~K and $C = 7.71$ emu/mol$\cdot$K.
The effective moment, $\mu_{\mathrm{eff}}$, is calculated to be 7.85 $\mu_{\mathrm{B}}$/f.u. using $C = N_{\mathrm{A}}\mu_{\mathrm{eff}}^{2}/3k_{\mathrm{B}}$, where $N_{\mathrm{A}}$  is the Avogadro constant and $k_{\mathrm{B}}$ is the Boltzmann constant. The obtained  $\mu_{\mathrm{eff}}$ value is close to 7.94 $\mu_{\mathrm{B}}$, which is expected for a free Eu$^{2+}$ ion with the $4f^{7}$ electronic configuration.
Figure~\ref{fig:basic_properties}(d) shows the magnetization curves obtained at 2~K. When the applied magnetic field is parallel to the $c$ axis, the magnetization shows a rapid increase in the magnetic field range from 1.75 to 2.50 T. When the applied magnetic field was perpendicular to the $c$ axis, two anomalies were observed at 3.6 and 6.1 T, appearing as inflection points.
In both magnetic field directions, the magnetization did not reach 7$\mu_{\mathrm{B}}$ at 7 T, the highest magnetic field applied in the present study. This can be explained by the fact that the anomalies at higher magnetic fields, as shown in the previous studies~\cite{JPSJ-2025-Miyake}, were not observed in the present measurements.

\subsection{Determination of Magnetic Propagation Vector}
\begin{figure*}
	\includegraphics{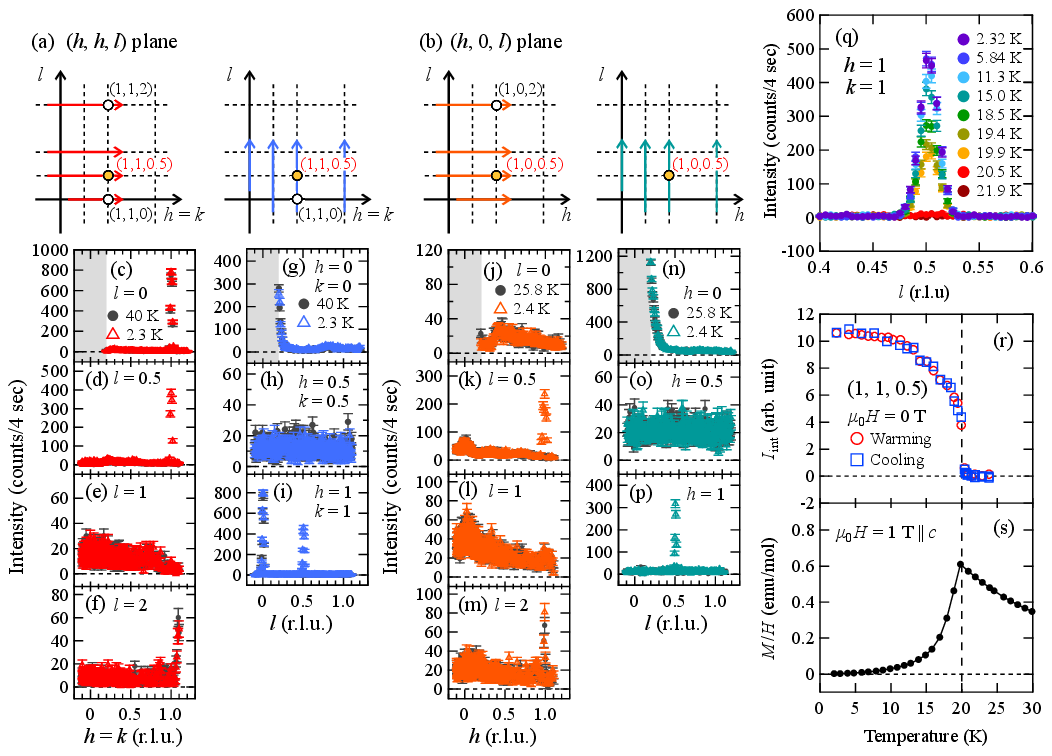}
	\caption{\label{fig_long_scan_HKL} (a,b) Scanning ranges of reciprocal space in the (a) $(h,h,l)$ and (b) $(h,0,l)$ planes. Observed nuclear Bragg reflections at (1,1,0), (1,1,2), and (1,0,2), and magnetic Bragg reflections at (1,1,0.5) and (1,0,0.5) are indicated by white- and yellow-filled circles, respectively. (c--i) Reciprocal space scans in the ($h,h,l$) plane at 40 K and 2.3 K. (j--p) Reciprocal space scans in the ($h,0,l$) plane at 25.8 K and 2.4 K. The gray-colored regions around (0, 0, 0) in (c), (g), (j) and (n) are not accessible due to the presence of the direct beam. (q) Temperature evolution of the magnetic Bragg peak at (1,1,0.5). (r,s) Temperature dependence of (r) the integrated intensity of the magnetic Bragg peak at (1,1,0.5) under the zero magnetic field and (s) magnetic susceptibility under a magnetic field of 1 T applied parallel to the $c$ axis.}
\end{figure*}

The search for magnetic Bragg peaks was performed using $h$ and $l$ scans in the $(h,h,l)$ and $(h,0,l)$ planes. The scanned reciprocal-space regions and the observed nuclear and magnetic reflections are schematically summarized in Figs.~\ref{fig_long_scan_HKL}(a) and~\ref{fig_long_scan_HKL}(b), and the corresponding scan profiles are shown in Figs.~\ref{fig_long_scan_HKL}(c--p). 

In the $(h,h,l)$ plane, nuclear reflections were observed at the (1,1,0) and (1,1,2) positions, whereas no reflection was identified at (1,1,1) above $T_{\mathrm{N}}$ [Figs.~\ref{fig_long_scan_HKL}(c),~\ref{fig_long_scan_HKL}(f) and~\ref{fig_long_scan_HKL}(i)].
These observations are consistent with the extinction rules for the tetragonal space group $P4/nmm$, which allow nuclear reflections only for even $l$ in the $(1,1,l)$ series.
In the antiferromagnetic state below $T_{\mathrm{N}}$, a magnetic reflection appeared at (1,1,0.5) [Figs.~\ref{fig_long_scan_HKL}(d) and \ref{fig_long_scan_HKL}(i)].
Similarly, in the $(h,0,l)$ plane, a nuclear reflection was observed at (1,0,2) above $T_{\mathrm{N}}$, whereas a magnetic reflection appeared at (1,0,0.5) below $T_{\mathrm{N}}$ [Figs.~\ref{fig_long_scan_HKL}(k), ~\ref{fig_long_scan_HKL}(m) and ~\ref{fig_long_scan_HKL}(p)].
These results indicate the fundamental magnetic propagation vector $\mathbf{q}=(0,0,1/2)$.
The line-scan profiles along the $(1,1,l)$ direction measured at various temperatures are shown in Fig.~\ref{fig_long_scan_HKL}(q).
From this data, the integrated intensity of the reflection at (1,1,0.5) was evaluated as a function of temperature, as shown in Fig.~\ref{fig_long_scan_HKL}(r).
The intensity gradually decreases with increasing temperature and disappears at $T_\mathrm{N} =$ 21 K, in good agreement with the peak in the magnetic susceptibility in a magnetic field applied along the $c$ axis [Figs.~\ref{fig_long_scan_HKL}(r) and ~\ref{fig_long_scan_HKL}(s)]. 
It is noted that no magnetic reflection was observed at (0,0,0.5) in the ($0,0,l$) line scan shown in Figs.~\ref{fig_long_scan_HKL}(g) and~\ref{fig_long_scan_HKL}(n). We also performed $\theta$-$2\theta$ scans at (0,0,2.5) and (0,0,3.5), and observed no discernible magnetic intensities.
In neutron scattering experiments, an intensity of a magnetic Bragg reflection is proportional to the square of Fourier-transformed magnetic moments projected onto the plane perpendicular to the scattering vector. 
Therefore, the absence of the magnetic reflection on the  ($0,0,l$) line, on which the scattering vector is parallel to the $c$ axis, indicates that the Eu moments are aligned along the $c$ axis, in agreement with the easy-axis behavior of the magnetic susceptibility.

\subsection{Magnetic Structure Analysis}
\begin{figure*}
	\includegraphics{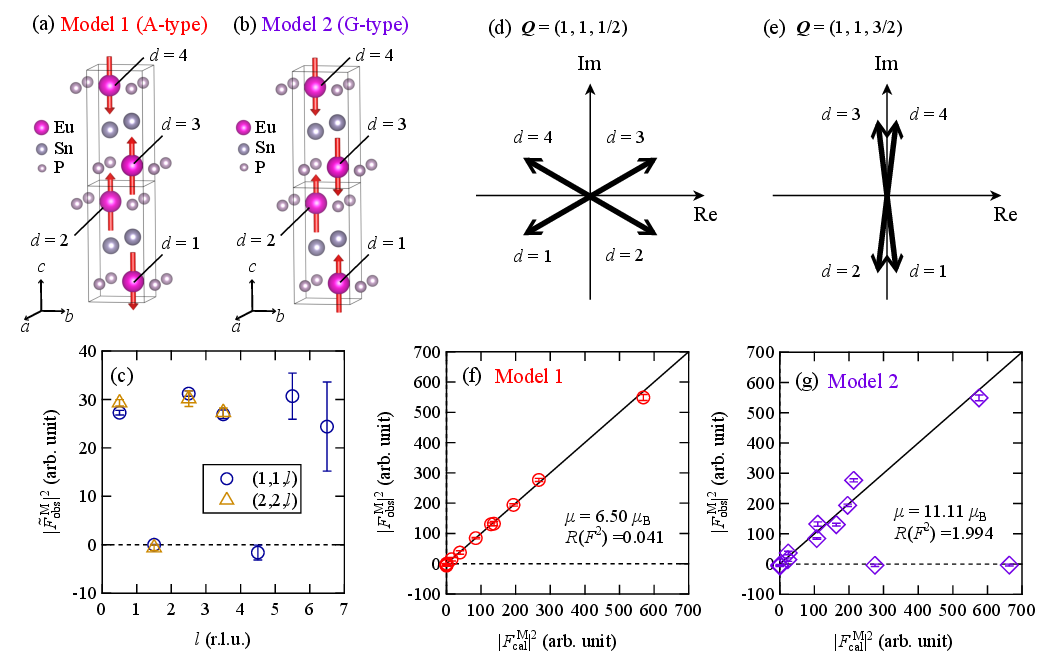}
	\caption{\label{fig_magnetic_structure} (a,b) Candidate magnetic structures characterized by a magnetic propagation vector $\mathbf{q} = (0, 0, 1/2)$, with magnetic moments aligned parallel to the $c$ axis. 
	The nearest-neighbor Eu moments are aligned (a) parallel in Model~1 (A-type) and (b) antiparallel in Model~2 (G-type). (a) and (b) were drawn using visualizing software VESTA~\cite{VESTA}. (c) The $l$-dependence of $|\tilde{F}^{\mathrm{M}}_{\mathrm{obs}}|^2 \equiv \frac{|F_{\mathrm{obs}}^{\mathrm{M}}|^{2}}{f^{\mathrm{Eu^{2+}}}(Q)^2[1-(\hat{\mathbf{Q}}\cdot\hat{\mathbf{c}})^2]}$ for the magnetic reflections at ($1,1,l$) and ($2,2,l$).
	(d,e) Phase factors $\exp(i\mathbf{Q}\cdot \mathbf{r}_{d})$ ($d = 1$–$4$) for (d) $\mathbf{Q}=(1,1,1/2)$ and (e) $\mathbf{Q}=(1,1,3/2)$. (f,g) Observed versus calculated squared magnetic structure factors, $|F^{\mathrm{M}}_{\mathrm{obs}}|^2$ versus $|F^{\mathrm{M}}_{\mathrm{cal}}|^2$, for (f) Model~1 and (g) Model~2. The solid line represents $|F^{\mathrm{M}}_{\mathrm{obs}}|^2 = |F^{\mathrm{M}}_{\mathrm{cal}}|^2$.}
\end{figure*}

Based on the above results, we can propose two candidate magnetic structures with the Eu moments aligned along the $c$ axis and the magnetic propagation vector $\mathbf{q}=(0,0,1/2)$.
In Model~1 (so-called A-type), the nearest-neighbor Eu moments are aligned ferromagnetically within the layer and antiferromagnetically along the stacking direction, whereas in Model~2 (so-called G-type), the nearest-neighbor Eu moments are aligned antiferromagnetically, as shown in Figs.~\ref{fig_magnetic_structure}(a) and~\ref{fig_magnetic_structure}(b), respectively. 
These magnetic structures are consistent with those proposed in the previous M\"{o}ssbauer study~\cite{SciRep-2025-Podgorska}.
To distinguish between these two candidate magnetic structures,  we compare the observed and calculated magnetic structure factors, $|F^{\mathrm{M}}_{\mathrm{obs}}|$ and $|F^{\mathrm{M}}_{\mathrm{cal}}|$, respectively.

Following the standard expression for neutron scattering~\cite{Squires}, the calculated magnetic structure factor is given by the sum of the magnetic scattering amplitudes from the Eu sites in the magnetic unit cell:
\begin{equation}
		F^{\mathrm{M}}_{\mathrm{cal}}(\mathbf{Q}) = \frac{1}{2}\sum_{d} \left(-\frac{\gamma r_{0}}{2}\right) {\mu}_{d}^{\perp} f^{\mathrm{Eu^{2+}}}(Q) \exp(i\mathbf{Q}\cdot \mathbf{r}_{d}) \exp(-W)
\end{equation}
where $(\gamma r_{0})/2 = 2.7$~fm is the prefactor relating the magnetic moment (in units of $\mu_{\mathrm{B}}$) to the magnetic scattering amplitude, $\mu_{d}^{\perp}$ is the component of the $d$-th Eu moment perpendicular to the scattering vector ($d=1$--4), $f^{\mathrm{Eu^{2+}}}(Q)$ is the magnetic form factor for Eu$^{2+}$, and $\mathbf{r}_{d}$ is the position vector of the $d$-th Eu site.
The Debye--Waller factor, $\exp(-W)$, depends only weakly on $\mathbf{Q}$ and is common to all the Eu sites.
For the present collinear magnetic structure with the moments aligned along the $c$ axis, the magnetic moment at the $d$-th Eu site can be written as
\begin{equation*}
	\boldsymbol{\mu}_{d}=\sigma_{d}\mu\hat{\mathbf{c}}
\end{equation*}
where $\sigma_d$ ($\pm1$) denotes the relative sign of the $d$-th Eu moment, $\mu$ is the magnitude of the ordered Eu moment, and $\hat{\mathbf c}$ is the unit vector parallel to the $c$ axis.
The perpendicular component is therefore given by
\begin{equation*}
	{\mu}_{d}^{\perp}=\sigma_{d}\mu\sqrt{1-(\hat{\mathbf{Q}}\cdot\hat{\mathbf{c}})^2}
	\end{equation*}
where $\hat{\mathbf{Q}}$ is the unit vector parallel to the scattering vector. 
Because the two candidate models differ only in the relative signs $\sigma_d$ of the four Eu moments (Table~\ref{tab:spin_config}), dividing the observed magnetic intensities by the common factors yields
\begin{equation}\label{eq:F2_f2cos2}
	\frac{|F_{\mathrm{obs}}^{\mathrm{M}}|^{2}}{ f^{\mathrm{Eu^{2+}}}(Q)^2[1-(\hat{\mathbf{Q}}\cdot\hat{\mathbf{c}})^2]} \propto \left| \sum_d \sigma_d \exp(i\mathbf{Q}\cdot \mathbf{r}_d) \right|^{2}.
\end{equation}
\begin{table}
	\caption{\label{tab:spin_config} Atomic coordinates from Ref.~\cite{JAllyoysCompd-2002-Payne} and the sign of $c$-axis component of magnetic moment, ${\sigma}_{d}$, of the $d$-th Eu$^{2+}$ ions.}
	\begin{ruledtabular}
		\begin{tabular}{cccccccc}
			$d$ & $x$ & $y$ & $z$ & ${\sigma}_{d}$ for Model 1 & ${\sigma}_{d}$ for Model 2 \\
			\hline
			1 & 0.75000 & 0.75000 & 0.16934 & -1 & 1 \\
			2 & 0.25000 & 0.25000 & 0.83066 & 1 & 1 \\
			3 & 0.75000 & 0.75000 & 1.16934 & 1 & -1 \\ 
			4 & 0.25000 & 0.25000 & 1.83066 & -1 & -1 \\ 
		\end{tabular}
	\end{ruledtabular}
\end{table}
In Fig.~\ref{fig_magnetic_structure}(c), we show the $l$ dependence of the quantity on the left-hand side of Eq.~(\ref{eq:F2_f2cos2}), which exhibits minima at $l = 3/2$ and $9/2$.
To understand this characteristic $l$-dependence,  we consider the right-hand side of Eq.~(\ref{eq:F2_f2cos2}).
Figures~\ref{fig_magnetic_structure}(d) and~\ref{fig_magnetic_structure}(e) illustrate the phase factors $\exp(i\mathbf{Q}\cdot\mathbf{r}_d)$ for $d=1$--$4$ on the complex plane at $\mathbf{Q}=(1,1,1/2)$ and $\mathbf{Q}=(1,1,3/2)$, respectively.
Since the $z$-coordinates of the four Eu sites, as listed in Table~\ref{tab:spin_config}, are $z_{0}$, $1-z_{0}$, $1+z_{0}$, and $2-z_{0}$, respectively, with $z_{0}\approx 1/6$, $\mathbf{Q}\cdot \mathbf{r}_d$ is approximately $\pm \pi/2$ at $l=3/2$ [Fig.~\ref{fig_magnetic_structure}(e)]. 
Accordingly, the four terms $\sigma_d \exp(i\mathbf{Q}\cdot \mathbf{r}_d)$ for Model~1 nearly cancel one another at $l=3/2$, whereas they do not cancel out at $l=1/2$, where $\mathbf{Q}\cdot \mathbf{r}_{d} \approx \pm \pi/6$ and $\pm 5\pi/6$ [Fig.~\ref{fig_magnetic_structure}(d)]. The suppressed intensity at $l=9/2$ is explained in the same manner.
These cancellations do not occur in Model~2, and the observed $l$-dependence therefore agrees better with Model~1 than with Model~2.

To quantitatively evaluate the validity of Model~1, least-squares refinements were performed for Models~1 and~2 using eight magnetic reflections. Here, the ordered Eu moment $\mu$ was used as the refinement parameter. For Model~1, the best-fit value was obtained to be $\mu = 6.50 \pm 0.03 \mu_{\mathrm{B}}$, which is reasonably close to the value expected for Eu$^{2+}$ with the $4f^{7}$ electronic configuration (see Appendix~\ref{appendix:structure_analysis} for the detailed analysis). 
As shown in Fig.~\ref{fig_magnetic_structure}(f), the observed and calculated squared magnetic structure factors show good agreement, with an $R(F^{2})$ factor of 0.041.
In contrast, Model~2 yields a much larger ordered moment of $\mu = 11.11 \pm 0.06  \mu_{\mathrm{B}}$ and a larger $R(F^{2})$ factor of 1.994 [Fig.~\ref{fig_magnetic_structure}(g)], indicating poor agreement with the experimental results.

We therefore conclude that the zero-field magnetic structure corresponds to Model~1, namely A-type antiferromagnetic structure in which Eu$^{2+}$ moments are aligned ferromagnetically within each layer and antiferromagnetically along the stacking direction.

\section{Discussion}
The neutron diffraction results revealed that the zero-field ground state of EuSnP is a collinear antiferromagnet with an easy axis along the $c$ axis, characterized by ferromagnetic intralayer and antiferromagnetic interlayer couplings of Eu$^{2+}$ moments [Fig.~\ref{fig_magnetic_structure}(a)]. 

With this result in mind, we revisit the magnetization curve in the magnetic field parallel to the $c$ axis~[Fig.~\ref{fig:basic_properties}(d)], where the abrupt increase in $M$ from 1.75 to 2.50 T is reminiscent of the  spin-flip transition in the typical A-type antiferromagnet FeCl$_{2}$, a van der Waals layered material~\cite{JPSJ-1964-Ito,JPSJ-1967-Jacobs,PRB-2023-Xu}.
From the structural viewpoint~[Fig.~\ref{fig:basic_properties}(a)], EuSnP may appear to be a quasi-two-dimensional magnet, similar to FeCl$_{2}$, in which the ferromagnetically aligned layers are coupled via weak antiferromagnetic interlayer interactions.

However, a key discrepancy from FeCl$_{2}$ lies in the paramagnetic susceptibility.
In general, two dimensionality of the system suppress $T_{\mathrm{N}}$.
When evaluated using the frustration parameter, $f = |\Theta_{\mathrm{W}}|/T_{\mathrm{N}}$, $f \sim 1$ for three-dimensional spin systems, whereas $f > 1$ for two-dimensional spin systems.
Indeed, for FeCl$_{2}$, $f \sim$ 48--84$/23.5\sim$ 2--4~~\cite{PR-1959-Wilkinson,PRB-1972-Yelon,PRB-2023-Xu} indicates a strong suppression of $T_{\mathrm{N}}$ due to low-dimensional fluctuations.
In contrast, for EuSnP, $f = 16/21 \sim  0.8$ which is close to or even below unity, suggesting that the spin system is three-dimensional.
It is also notable that the magnetic susceptibility follows the Curie-Weiss law down to 50 K, corresponding to approximately 2.5$T_{\mathrm{N}}$. 
This behavior is distinct from that of FeCl$_{2}$, in which the susceptibility deviates from Curie-Weiss relation at a temperature as high as 220 K~\cite{PRB-2023-Xu}, roughly 10 times of  $T_{\mathrm{N}}$,  reflecting the developing intralayer ferromagnetic fluctuations.

The validity of the mean-field description in the paramagnetic state of EuSnP discussed above suggests that the interlayer interaction is sufficiently strong for the system to be regarded as three-dimensional.
Long-range interactions, such as the Ruderman-Kittel-Kasuya-Yosida (RKKY) interaction between Eu moments mediated by conduction electrons, are likely responsible for this three-dimensional character. 
The ARPES studies reported that the Sn and P conduction electrons weakly couple to the magnetically ordered Eu moments, arguing that the complex geometry of the Fermi surfaces allows for sufficient coupling of Eu moments~\cite{PRMater-2024-Sprague}.
The  presence of multiple metamagnetic transitions~\cite{PhysicaB-2006-Fujiwara,SciRep-2025-Podgorska,JPSJ-2020-Onuki} also suggests the possible emergence of noncollinear spin structures, which may arise from the field-induced modulation of RKKY interaction. 
Further investigations such as neutron scattering experiments under magnetic field, would be crucial for identifying the detailed magnetic structure.

Finally, we discuss the possibility that Eu--P orbital hybridization contributes to the enhancement of the intralayer ferromagnetic couplings.
According to previous high-pressure structural studies~\cite{DaltonTrans-2019-Gui}, Eu--P bonding is significantly strengthened under pressure, whereas Sn--P and Sn--Sn bonding remain nearly unchanged or slightly weakened.
These features imply that Eu-P hybridization plays an important role in the increase of $T_{\mathrm{N}}$ under pressure.
Here we compare the present system with semiconducting ferromagnet EuO~\cite{PRL-2009-Miyazaki}.
In EuO, the ferromagnetism of 4f moments in Eu$^{2+}$ arises from indirect exchange interactions via Eu 5d orbitals and superexchange interactions mediated by O 2p orbitals~\cite{PRL-2009-Miyazaki}.
In the case of EuSnP, the band calculations show finite Eu 4f-P 3p hybridization below the Fermi energy while the Fermi surface primarily consists of Sn 5p and P 3p orbitals~\cite{PRMater-2024-Sprague}.
A similar combination of indirect exchange and superexchange-like interactions to that proposed for EuO may be present in EuSnP.
A detailed consideration of the bonding geometry and the orbital character of the itinerant bands will be important for clarifying the microscopic origin of the competing exchange interactions underlying the A-type antiferromagnetic structure.

\section{Conclusions}
We have clarified the zero-field magnetic structure of EuSnP using single-crystal neutron diffraction.
We find that the ground state is an A-type antiferromagnet characterized by ferromagnetic intralayer and antiferromagnetic interlayer couplings between Eu$^{2+}$ moments.

The magnetic properties under external fields suggest the complexity of the underlying magnetic interactions, providing an opportunity to explore the interplay between itinerant electrons and localized moments in EuSnP.

\begin{acknowledgments}
We acknowledge Dr. R. Ishii for experimental support in crystal synthesis.
This work was supported by JSPS Grants in Aid for Scientific Research (Grants Nos. 24K00579, 23K25815, 24K17004 and 26K17084), the Asahi Glass Foundation, MST Foundation and a grant by the Noguchi institute.
The neutron experiment at  5G-PONTA was carried out by the JRR-3 general user program managed by the Institute for Solid State Physics, the University of Tokyo (proposal No. 23521). 
A part of the magnetization measurements were performed using facilities of the Cryogenic Research Center, the University of Tokyo.
\end{acknowledgments}

\appendix
\section{Refinement of the Scale Factor and Magnetic Moment}\label{appendix:structure_analysis}
In the present work, the scale factor between the observed and calculated structure factors was determined by least-squares fitting of the calculated nuclear structure factors $|F^{\mathrm{N}}_{\mathrm{cal}}|$ to the observed structure factors $|F^{\mathrm{N}}_{\mathrm{obs}}|$ at 2~K.
The nuclear reflections were measured by $\theta$--2$\theta$ scans, and the $|F^{\mathrm{N}}_{\mathrm{obs}}|$ values were obtained from the integrated intensities after correcting for neutron absorption by the constituent elements.
The $|F^{\mathrm{N}}_{\mathrm{cal}}|$ values were calculated using the crystal structure reported in Ref.~\cite{JAllyoysCompd-2002-Payne} without further refinement of the atomic positions.
In this analysis, the secondary extinction correction of Ref.~\cite{ActaCryst-1974-Becker} was applied.

\begin{figure}
	\includegraphics{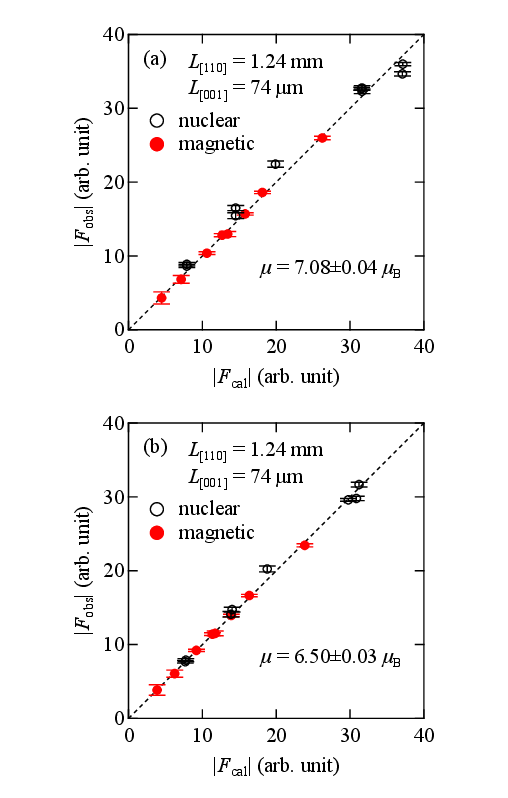}
	\caption{\label{fig:appendix_FcalFobs}  Comparison of the observed and calculated
		structure factors for nuclear and magnetic diffraction (see the text in Appendix~\ref{appendix:structure_analysis} for details).}
\end{figure}

Figure~\ref{fig:appendix_FcalFobs}(a) shows the best-fit result obtained using ten nuclear reflections together with the result of least-squares refinement of eight magnetic reflections, yielding $\mu = 7.08\pm 0.04 \mu_{\mathrm{B}}$.
Considering that all the magnetic reflections measured in the present work give relatively small $|F^{\mathrm{M}}_{\mathrm{cal}}|$ values, we excluded the nuclear reflections at (1,1,0) and (2,2,0), which have the largest $|F^{\mathrm{N}}_{\mathrm{cal}}|$ values, and performed the same analysis, yielding $\mu = 6.50\pm 0.03 \mu_{\mathrm{B}}$ [Fig.~\ref{fig:appendix_FcalFobs}(b)].
The values of $\mu$ obtained from the two analyses, 7.08 $\mu_{\mathrm{B}}$ and 6.50 $\mu_{\mathrm{B}}$, are both close to 7$\mu_{\mathrm{B}}$, consistent with the $4f^{7}$ electronic configuration of Eu$^{2+}$.
For a more precise determination of $\mu$, a more detailed analysis based on a larger number of reflection points is required.

In the above analysis, the absorption correction was applied by assuming that the sample has the rectangular shape with the dimensions of 1.24 mm $\times$ 2.00 mm $\times$ 74 $\mu$m along the [110], [1-10] and [001] directions, respectively. 
In the measurements on the ($h,h,l$) scattering plane, the lengths along the [110] and [001] directions ($L_{[110]}$ and $L_{[001]}$) are relevant to the absorption correction. 
To examine the effect of possible errors in measuring these lengths, we performed the same analysis with varying the length along the [001] direction by $\pm$10\% from the measured value ($L_{[001]}=$74~$\mu$m), i.e., 81.4 $\mu$m and 66.6 $\mu$m, as well as using the mean sample width $L_{[110]}=$0.992 mm.
Nevertheless, the best-fit $\mu$ value differs by less than 1\% from that obtained using $L_{[110]}=1.24$~mm and $L_{[001]} = $74~$\mu$m.

In the main text, we present the result shown in Fig.~\ref{fig:appendix_FcalFobs}(b).
Using the obtained parameters, the $R(F)$-factor was 0.02693 for the nuclear reflections and 0.01282 for the magnetic reflections.


%

\end{document}